\documentclass{article}
\usepackage[T1]{fontenc}
\usepackage[utf8]{inputenc}
\usepackage{ismir}
\usepackage{amsmath,cite,url}
\usepackage{graphicx}
\usepackage{color}
\usepackage{booktabs}
\usepackage{multirow}
\usepackage{amssymb}
\usepackage{siunitx}
\usepackage{makecell}
\usepackage[dvipsnames]{xcolor}
\usepackage{colortbl}
\definecolor{ours}{RGB}{219, 234, 254}

\title{Local Multimodal Music Alignment \\ from Global Supervision}

\multauthor
  {Irmak Bukey$^1$ \hspace{0.5cm} Zachary Novack$^2$ \hspace{0.5cm} Jongmin Jung$^3$ \hspace{0.5cm} Dasaem Jeong$^4$ \hspace{0.5cm} Chris Donahue$^1$}
  {
  $^1$ Carnegie Mellon University
  $^2$ University of California San Diego
  $^3$ Neutune
  $^4$ Sogang University \\
  }

\def\authorname{I. Bukey, Z. Novack, J. Jung, D. Jeong, and C. Donahue}

\usepackage[bookmarks=false,pdfauthor={\authorname},pdfsubject={\pdfsubject},hidelinks]{hyperref}

\begin{document}

\newcommand{\segment}{segment}
\newcommand{\segments}{segments}
\newcommand{\atom}{frame}
\newcommand{\atoms}{frames}
\newcommand{\ib}[1]{{\color{blue}{{[Irmak: #1]}}}}
\newcommand{\cd}[1]{{\color{magenta}{{[Chris: #1]}}}}
\newcommand{\zn}[1]{{\color{ForestGreen}{{[Zack: #1]}}}}
\newcommand{\todo}[1]{{\color{red}{{[TODO: #1]}}}}
\newcommand{\da}[1]{{\color{cyan}{{[Dasaem: #1]}}}}
\newcommand{\batchsim}{\mathbf{S}}
\newcommand{\framesim}{\mathbf{F}}
\maketitle
\begin{abstract}
Understanding music requires understanding localized relationships across data modalities, e.g., 
how time in performance audio maps onto position in a score image.
Yet supervision for such \emph{local} correspondences is difficult to obtain---in practice, we often only have access to coarser \emph{global} supervision like paired \segments{} of audio and images. 
To address this gap, we propose \textbf{FuSiLi} (\textbf{Fu}sed \textbf{Si}nkhorn-\textbf{L}ocalized S\textbf{i}milarity), a similarity score for multimodal contrastive learning operating directly on local image patch and audio frame features via Sinkhorn-based soft alignment. We show that FuSiLi 
(i)~effectively learns local relationships,
(ii)~requires only global supervision, and
(iii)~retains the global alignment capabilities of conventional contrastive approaches. 
We fine-tune pre-trained CLIP and CLAP encoders on pairs of raw sheet music images and audio using a hybrid contrastive objective combining FuSiLi with conventional global similarity.
We evaluate on cross-modal retrieval and frame-level alignment tasks against a range of global and local baselines, showing that our approach outperforms them on local alignment while remaining competitive on retrieval.
\end{abstract}
\section{Introduction}

Many multimodal music understanding tasks involve localized correspondences between data modalities~\cite{benetos2018automatic, calvo2020understanding, orio2003score}.
For example, we may want to understand how performance audio aligns to an image of the corresponding sheet music. 
Ideally, supervision for this task would involve two types of ground truth information:
(1)~\emph{global} supervision pairing the right audio with the right image, and
(2)~a \emph{local} alignment specifying a mapping between time in the audio and pixels in the image. 
If we could collect both forms of supervision at scale, we could learn to predict the local alignment for a new pair of inputs in a fully supervised fashion. 
However,
it is prohibitively expensive to collect local alignments at scale in practice, as it requires significant manual effort from expert annotators~\cite{bukey2024just, feffer2022assistive}.
Compared to the expense of collecting local alignments, it is much cheaper to collect global supervision in the form of audio-image pairs that correspond to the same piece. 
Such supervision scales naturally from numerous sources, as 
IMSLP 
provides 
pairs at the piece-level, 
while datasets such as YTSV~\cite{jung2025unified} offer pairs on more granular \emph{\segments} (a few measures of music). 
This broad availability motivates our key research question: \textbf{can we learn local alignments across modalities from global supervision alone?}

The dominant approach for leveraging multimodal global supervision at scale is contrastive learning.
Models like CLIP~\cite{radford2021learning} and CLAP~\cite{elizalde2023clap, wu2023large} have demonstrated that contrastive objectives applied to globally paired data can yield powerful multimodal representations. However, we show that this approach is not effective at capturing local alignments, achieving only $16$\% top-1 accuracy on a \emph{\atom-level} alignment task (here, aligning time frames of audio to pixel patches in the score image). Other specialized approaches attempt to learn localized alignments but carry significant drawbacks: some need explicit local supervision~\cite{wu2025flam}, while others are prohibitively expensive for retrieval~\cite{jung2025unified}, requiring
$|Q| \times |R|$ forward passes (query/retrieval sets $Q$/$R$), compared to the $|Q|+|R|$ passes for contrastive models.

To address this gap, 
we propose a new inductive bias for multimodal contrastive learning. Existing contrastive approaches follow a \emph{pool-then-fuse} paradigm: local features from unimodal encoders are first pooled into global vectors and similarity is computed as a single dot product between these pooled vectors. We instead propose a \emph{fuse-then-pool} paradigm: we compute pairwise cosine similarity directly between \atom-level features, weight the resulting matrix using a Sinkhorn-derived soft alignment to capture \atom-level correspondences between modalities, and only then pool to produce the scalar similarity score required by the contrastive objective. We refer to this approach as \textbf{FuSiLi} (\textbf{Fu}sed \textbf{Si}nkhorn-\textbf{L}ocalized S\textbf{i}milarity). This reformulation allows fine-grained local alignment to emerge from global supervision only, while preserving the 
asymptotic 
inference efficiency of the standard contrastive models.

We implement FuSiLi by fine-tuning pre-trained image and audio encoders~\cite{radford2021learning, wu2023large} on \segment-level pairs of sheet music images and audio using the standard InfoNCE objective with our proposed similarity score formulation. FuSiLi achieves $30$\% top-1 accuracy on frame-level alignment, a substantial gain over the $16$\% of standard contrastive baselines, while preserving strong cross-modal retrieval performances (FuSiLi: $43.5$\% vs. Baseline: $43.8$\% average MRR across datasets). Furthermore, we demonstrate zero-shot capability on real-world tasks such as mapping a position in the image to its corresponding audio timestamp in the full recording,
where our method achieves over $10\times$ improvement over the baseline ($13.91\%$ vs $1.33\%$). 
We provide qualitative examples demonstrating the performance of our model\footnote{Examples: \url{https://irmakbky.github.io/fusili/}} and our code\footnote{Code: \url{https://github.com/irmakbky/fusili}}.

\begin{figure*}[h!]
    \centering
    \includegraphics[width=\linewidth]{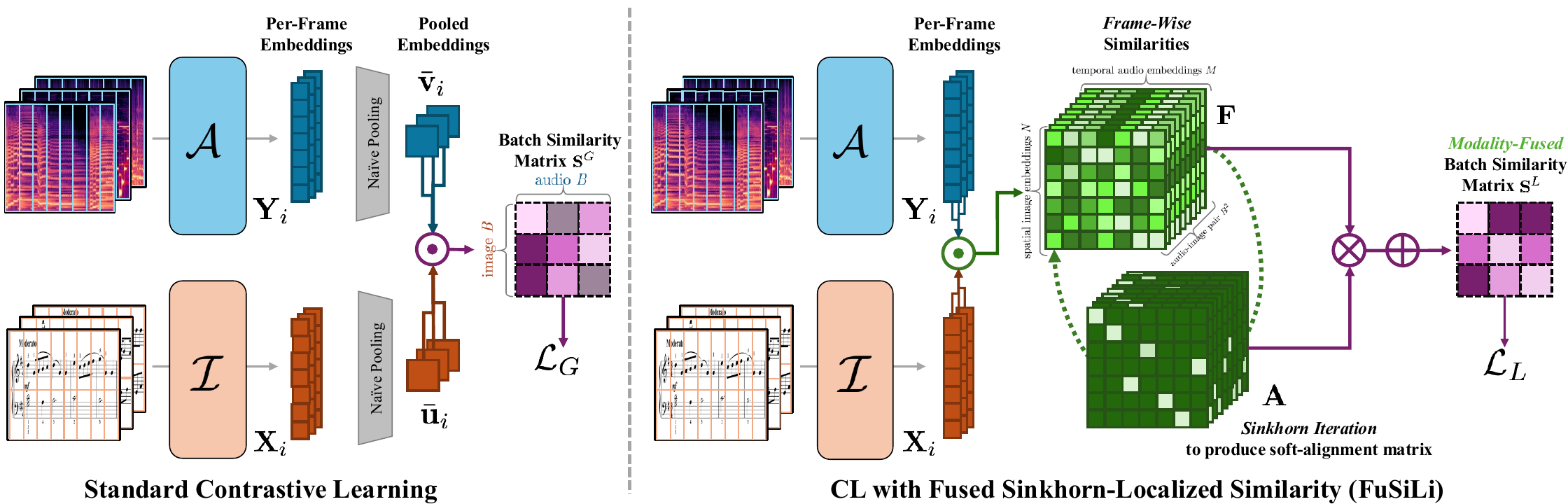}
    \caption{Pipeline of Standard contrastive learning vs. \textbf{FuSiLi}. Standard approaches perform the batch-wise similarity calculation on pre-pooled representations, removing any capacity for explicit cross-modal locality. With our \emph{fuse-then-pool} FuSiLi, we calculate frame-wise similarity matrices $\framesim_{ij}$ for each batch pairing and perform Sinkhorn iteration to enforce one-to-one correspondences at the frame level, and \emph{then} pool into global embeddings which now encode local information.}
    \label{fig:fig1}
    \vspace{-1em}
\end{figure*}
\section{Related Work}
\label{sec:related_work}

\textbf{Audio-score alignment.}
The task of aligning audio performances to musical scores has a long history in music information, with classical \emph{score following} approaches dating back to early work on online real-time algorithms~\cite{dannenberg1984line, vercoe1984synthetic} and later extending to 
Dynamic Time Warping~\cite{dixon2005line}  
and Hidden Markov Models~\cite{cont2006realtime}.
These works typically operate on symbolic representations rather than raw score images, and target online performance tracking rather than the offline cross-modal alignment.
End-to-end approaches that learn directly from sheet music images and audio have been explored 
~\cite{dorfer2017learning, dorfer2018learning, balke2019learning, henkel2021real, bukey2024just, carvalho2023towards, carvalho2023self};
however, these works either rely on fine-grained local supervision or focus on global alignment/retrieval, making them less directly comparable to our setting. A parallel line of work frames cross-modal music translation as a sequence-to-sequence problem~\cite{jung2025unified},
but requires $|Q| \times |R|$ forward passes for query/retrieval sets $Q$ and $R$. 
Our work builds on the contrastive paradigm for efficient $|Q|+|R|$ inference while addressing its limitation of not capturing fine-grained local structure.

\textbf{Multimodal contrastive learning.}
Models such as CLIP~\cite{radford2021learning} and CLAP~\cite{elizalde2023clap, wu2023large} have demonstrated that encoders trained with InfoNCE~\cite{oord2018representation} on large paired datasets yield powerful transferable representations.
Music-specific variants further adapt these ideas to musical content~\cite{wu2025clamp, manco2022contrastive, wu2025collap}. All these methods aggregate local features into global embeddings before computing similarity, which we show is insufficient for learning frame-level correspondences.

\textbf{Localized contrastive learning.}
Several works augment contrastive frameworks with objectives encouraging local alignment. In the language-audio domain, work relies on explicit frame-level annotations as supervision for local objectives~\cite{wu2025flam, li2026finelap, primus2025tacos}, besides~\cite{li2024advancing} which uses weakly-aligned data for multi-grained audio-language alignment but uses an attention-based mechanism for local representations.
In vision, more works employ attention-based methods~\cite{zeng2021contrastive, anonto2025align} or rely on explicit region-level supervision~\cite{chen2024contrastive}. 
 Broadly, prior work has found that patch-level features in globally trained models carry semantically meaningful local information without explicit supervision during training~\cite{gandelsman2023interpreting, bousselham2024grounding}. 
Our approach differs from all of the above in that we modify the \emph{similarity function} itself to encourage frame-level alignment to emerge from global segment-level supervision alone.

\textbf{Optimal transport 
in representation learning.}
The Sinkhorn algorithm offers an efficient entropic approximation to optimal transport~\cite{cuturi2013sinkhorn} and has seen growing use across representation learning settings~\cite{caron2020unsupervised, emon2025whisq, chen2024your}. 
Our use of Sinkhorn differs from prior work:
rather than applying it at the batch level to reweight sample pairs, we apply it at the instance level to compute a soft alignment between image patches and audio frames within each pair.
\section{Method}

\subsection{Multimodal Contrastive Learning Preliminaries}
\label{subsec:prelims}

Multimodal contrastive learning frameworks align representations from different modalities by optimizing a similarity-based objective over paired data. Most commonly, these frameworks use the InfoNCE objective~\cite{oord2018representation} which is defined over a batch of $B$ pairs as:
\begin{equation}
    \mathcal{L}_{\batchsim} = -\frac{1}{B} \sum_{i=1}^{B} \log \frac{\exp(\batchsim_{ii} / \tau)}{\sum_{j=1}^{B} \exp(\batchsim_{ij} / \tau)}
\end{equation}
where $\tau$ is the temperature hyperparameter, and ${\batchsim \in \mathbb[-1, 1]^{B \times B}}$ is a $B \times B$ matrix of similarities, where $\batchsim_{ij}$ denotes the similarity between sample $i$ from one modality and sample $j$ from the other.
In practice, CLIP-style models compute this as a bidirectional loss,
treating samples from one modality as queries against all samples from the other modality, and vice versa, averaging the two terms. We follow this convention throughout.
This objective minimizes the distance between matching pairs (positive pairs) and pushes apart the mismatched (negative) pairs in the embedding space. 
\textbf{The key distinction between the conventional approach and our proposed approach is how the similarity matrix $\batchsim$ is calculated}, which we detail below and illustrate in Figure~\ref{fig:fig1}. 

Under the hood, 
CLIP-style contrastive learning implicitly relies on the extraction of localized features from individual modalities. 
For a batch of $B$ pairs of sheet music images and audio clips, modality-specific encoders extract image features ${\mathbf{X}_i \in \mathbb{R}^{N \times d_{\text{img}}}}$ and audio features ${\mathbf{Y}_i \in \mathbb{R}^{M \times d_{\text{audio}}}}$, where $N$ and $M$ denote the number of image patches (a square group of pixels, e.g., $7 \times 7$), and audio frames (short segments of time, e.g., $78.1$ms), and $i \in \{1, \ldots, B\}$. For simplicity, henceforth we refer to both image patches and audio frames as \emph{\atoms}, and their alignment as \atom-level. 

\subsection{Conventional Global Contrastive Scores ($\batchsim^G$)}
\label{subsec:globalscores}
The conventional approach in contrastive learning 
computes similarity between globally pooled embeddings by collapsing 2D features into 1D embeddings vectors:
\begin{equation}
    \bar{\mathbf{u}}_i = \text{Proj}^G_{\text{img}}(\text{Pool}_{\text{img}}(\mathbf{X}_i)), \text{ }
    \bar{\mathbf{v}}_i = \text{Proj}^G_{\text{aud}}(\text{Pool}_{\text{aud}}(\mathbf{Y}_i)),
\end{equation}
where $\bar{\mathbf{u}}_i, \bar{\mathbf{v}}_i \in \mathbb{R}^d$,
$\text{Pool}_{\text{img}}:~\mathbb{R}^{N \times d_{\text{img}}} \rightarrow \mathbb{R}^{d_{\text{img}}}$ aggregates image patches into a single vector (e.g.,~mean pooling or \texttt{[CLS]} token), $\text{Pool}_{\text{aud}}:~\mathbb{R}^{N \times d_{\text{audio}}} \rightarrow \mathbb{R}^{d_{\text{audio}}}$ aggregates audio frames into a single vector (e.g.,~mean pooling or token-semantic mapping), and $\text{Proj}_{\text{img}}^G: \mathbb{R}^{d_{\text{img}}} \rightarrow \mathbb{R}^{d}$ and $\text{Proj}_{\text{aud}}^G: \mathbb{R}^{d_{\text{audio}}} \rightarrow \mathbb{R}^{d}$ are learned projections into the shared $d$-dimensional embedding space.
The similarity score between image $i$ and audio $j$ is then the cosine similarity between their global embeddings:
\begin{equation}
    \batchsim_{ij}^{G} = \frac{\bar{\mathbf{u}}_i^\top \bar{\mathbf{v}}_j}{\|\bar{\mathbf{u}}_i\| \|\bar{\mathbf{v}}_j\|}
    \label{eq:batchsim_g}
\end{equation}
$\batchsim_{ij}^{G}$ reflects the overall similarity between the two samples, but fails to fully capture frame-level correspondences. We denote the global contrastive loss as $\mathcal{L}_G = \mathcal{L}_{\batchsim^G}$.

\subsection{Local Contrastive Scores ($\batchsim^L$) with FuSiLi}
\label{subsec:localloss}

Instead of pooling before computing similarity, we propose operating \emph{directly} on \atom-level 
features and pooling only after local correspondences have been captured. We first project the local image and audio features $\mathbf{X}_i, \mathbf{Y}_i$ into a shared $d$-dimensional  space via modality-specific linear projection layers. These projections preserve the spatial and temporal structure of the 
local features.

\subsubsection{Soft Alignment via Sinkhorn}
\label{subsec:sinkhorn_align}
For each pair $(i, j)$, we compute a \atom-wise alignment matrix $\framesim_{ij} \in [-1, 1]^{N \times M}$ (distinct from ${\batchsim \in [-1, 1]^{B \times B}}$). 
Each entry $(n, m)$ in $\framesim_{ij}$ is:
\begin{equation}
    [\framesim_{ij}]_{nm} = \frac{\mathbf{x}_{i,n}^\top \mathbf{y}_{j,m}}{\|\mathbf{x}_{i,n}\| \|\mathbf{y}_{j,m}\|}
\end{equation}
where $\mathbf{x}_{i,n}$ is the $n$-th \atom{} of 
$\mathbf{X}_i$ projected into the shared embedding space,
and $\mathbf{y}_{j,m}$ is the $m$-th \atom{} of
$\mathbf{Y}_j$ projected into the shared embedding space.
Intuitively, 
in $\framesim_{ii}$ (a matched pair where $i = j$), entries corresponding to spatiotemporally aligned \atoms{} should be high. 

Next, we apply the Sinkhorn algorithm. The Sinkhorn algorithm is traditionally used to solve Optimal Transport problems~\cite{cuturi2013sinkhorn} by transforming a cost or similarity matrix into a doubly stochastic matrix (i.e.~that each row and column sums to 1). 
In our method, it is used to convert the \atom-level similarity matrix $\framesim_{ij}$ into a soft alignment matrix $\mathbf{A}_{ij}$ between the image and audio \atoms.
While a natural alternative would be to use the differentiable alignment cost from SoftDTW~\cite{cuturi2017soft}, this assumes monotonic sequence alignment.
This assumption does not hold for sheet music, where line breaks and vertical structure mean image patches do not follow a simple temporal order (Figure~\ref{fig:fig2}). Sinkhorn makes no such assumption, instead learning a flexible frame-level correspondence.

We initialize $\mathbf{P}_{ij}^{(0)} = \exp(\framesim_{ij} / \epsilon)$, where $\epsilon$ is a temperature hyperparameter that controls the sharpness of the alignment. We then perform $K$ iterations of alternating row and column normalization
\begin{equation}
\mathbf{P}^{(k+1)}_{ij} = \operatorname{softmax}_{c}(\operatorname{softmax}_{r}(\mathbf{P}^{(k)}_{ij}))
\end{equation}
where $\operatorname{softmax}_{r}$ and $\operatorname{softmax}_{c}$ denote row- and column-wise softmax.
The resulting soft-alignment matrix $\mathbf{A}_{ij} = \mathbf{P}^{(K)}_{ij} \in
\mathbb{R}_{> 0}^{N \times M}$
captures \atom-level correspondences between image and audio.
Unlike the softmax operation applied in a single direction in standard attention, iterating this normalization process in both directions enforces mutual constraints, ensuring that similarities are distributed across both image and audio frames simultaneously.

\subsubsection{Fusing Local Alignment into $\batchsim^L$}
\label{subsec:fusili}

Using the soft-alignment matrix $\mathbf{A}_{ij}$, we compute a scalar local similarity score as the sum of the element-wise product between the alignment matrix and the similarity matrix:
\begin{equation}
    \batchsim_{ij}^{L} = \sum_{n=1}^{N} \sum_{m=1}^{M} [\mathbf{A}_{ij}]_{nm} \cdot [\framesim_{ij}]_{nm}
    \label{eq:local_sim}
\end{equation}
Intuitively, $\batchsim_{ij}^{L}$ is high when frames that have high cosine similarity also receive high alignment probability after the Sinkhorn iterations. As a result, \textbf{$\mathbf{S}_{ij}^L$ captures the most semantically relevant frame-level correspondences, using the Sinkhorn alignment as a soft weighting over the similarity matrix rather than treating all frame pairs equally}. An overview of our method can be seen in Fig.~\ref{fig:fig1}.
The local contrastive loss is defined as $\mathcal{L}_L = \mathcal{L}_{\batchsim^L}$

\subsection{Hybrid Training Objective}
\label{subsec:dualobjective}

Our primary training objective uses the InfoNCE loss with the FuSiLi score, $\mathcal{L}_L$, which requires only segment-level supervision while implicitly learning frame-level alignment. We additionally investigate whether combining $\mathcal{L}_L$ with the global contrastive loss $\mathcal{L}_G$ further improves performance, particularly on retrieval tasks where global semantic similarity can be efficiently encoded by pooled global embeddings. We defined this hybrid loss as:
\begin{equation}
    \mathcal{L} = \alpha \mathcal{L}_{L} + (1 - \alpha) \mathcal{L}_{G}
\end{equation}
where $\alpha \in [0, 1]$ is a hyperparameter balancing the two loss terms. We treat $\alpha$ as an experimental variable, studying both the case where training relies on $\mathcal{L}_L$ alone ($\alpha = 1$) and where both objectives contribute ($0 < \alpha < 1$). 

\subsection{Inference}
\label{subsec:inference}

For inference, global retrieval ranks candidates by $\batchsim^G$ as defined in Equation~\ref{eq:batchsim_g} for models trained with a global objective ($\alpha < 1$), or by the summed pairwise cosine similarity $\sum_{n,m} [\framesim_{ij}]_{nm}$ for models trained with $\mathcal{L}_L$ only ($\alpha=1$).
For local similarity, FuSiLi offers a desirable property at inference time: despite being trained with the iterative Sinkhorn algorithm, it does not require it for inference. Direct cosine similarity between projected \atom-level embeddings is sufficient to produce \atom-level alignments. We empirically verify that this does not affect alignment quality, while substantially reducing inference cost ($37$ms $\to$ $5$ms). We attribute this to backpropagating through Sinkhorn iteration during training, which may allow the alignment structure to be encoded in the initial local embeddings, making cosine similarity sufficient.
\section{Experiments}

\subsection{Data}
\label{subsec:datasets}
We derive synthetic performance audio and engraved score images from MusicXML via software rendering, yielding globally paired image--audio segments without requiring manual annotations. While our MusicXML rendering pipeline could theoretically provide local alignments for the paired image/audio segments, the software tools we use do not provide this information in practice. 

We train on a $\approx$40K unique piece subset of PDMX \cite{long2025pdmx}, a large-scale dataset of public domain MusicXML scores.
For each piece, we extract 8-measure segments with 50\% overlap, a window size chosen to correspond to approximately 20 seconds of audio.
Each segment is rendered into a $224\times224$ image and synthesized into a 20-second audio clip, yielding approximately 345k image-audio pairs in total of which 324k are used for training and the rest for validation and testing.
For the test set, we evaluate global retrieval only, using 406 segments per piece.

For localized evaluation, we use MSMD~\cite{dorfer2018learning}, a synthetic dataset providing ground-truth note-level alignments between score images and audio. We project these
note-level alignments into frame-level correspondences by mapping each note's onset--offset interval to the corresponding audio frames and its pixel location to the corresponding image patch index.
We treat each system in the score as a segment, discarding those exceeding 20 seconds.
All systems from the full train and test splits (423 pieces, 5337 segments) are used for local evaluation (validation split is used to track validation metrics during training); global retrieval uses one segment per piece (423 segments).

For out-of-domain evaluation, we use the string quartet subset of YTSV~\cite{jung2025unified}, a dataset of paired score images and audio performances, provided as segments of a few measures each from YouTube score videos.
Unlike the synthetic data in PDMX, this more challenging dataset contains expressive human performance audio aligned to sheet music image scans.
We retain 542 segments from distinct pieces after discarding those exceeding 20 seconds.

\subsection{Modeling Details}
\label{subsec:impl}

We fine-tune the CLIP ViT image encoder 
~\cite{ilharco_gabriel_2021_5143773}
and the LAION-CLAP HTSAT audio encoder 
~\cite{wu2023large,htsatke2022}.
Since the CLAP encoder natively accepts 10-second inputs, we split each clip into two 10-second segments, encode them separately, and concatenate the resulting frame embeddings. With downsampling, this yields 256 audio frames per clip.

\subsection{Baselines}
\label{subsec:baseline}
\label{subsec:contrastive_baseline}

Our primary baseline is a standard global contrastive model trained with $\mathcal{L}_G$ only ($\alpha = 0$, see Section~\ref{subsec:globalscores}). This corresponds to a conventional CLIP-style fine-tuning setup where paired image and audio inputs are pooled globally and the model is trained with InfoNCE. 

To situate our approach relative to an alternative framework where local alignments might emerge from global supervision, we include U-MusT~\cite{jung2025unified}, a sequence-to-sequence model for multimodal music translation. 
We use the checkpoint trained on YTSV-piano. 
Due to computational limitations, we were unable to retrain U-MusT on PDMX,
so this comparison should be interpreted 
as ``apples-to-oranges''.
Following the original work, we feed score images into the encoder and audio into the decoder, and extract
cross-attention matrices from decoder layers 7, 8, and 9. 
We average across layers and attention heads to obtain an $N \times M$ alignment matrix per pair, used as a 
substitute for the cosine similarity matrix in local evaluation.

We do not compare against score following systems as (1) we target raw score images, while score following systems typically require symbolic representations (e.g., MIDI), and (2) score following systems are tailored for online performance, while here we focus on offline scenarios.

\subsection{Evaluation}
\label{subsec:eval}

\textbf{Global retrieval.} We evaluate segment-level cross-modal retrieval in both the image-to-audio (I2A) and audio-to-image (A2I) directions on PDMX and MSMD (in-domain) and YTSV (out-of-domain) test sets. Similarities are computed as described in Section~\ref{subsec:inference}.
We report Recall@1 (R@1) and Mean Reciprocal Rank (MRR). The U-MusT baseline is excluded from this evaluation as it requires a separate forward pass per pair and would be impractical for global retrieval in real-world settings.

\textbf{Local alignment.} Frame-level alignment is evaluated on the MSMD eval set, which provides note-level correspondences between images and audio. Given a matched pair, we compute the pairwise cosine similarity matrix between image and audio embeddings, yielding an $N \times M$ matrix. For each audio frame, we predict the corresponding image patch by selecting the index with the highest cosine similarity and compare it to the ground-truth, reporting top-1 accuracy. For $\mathcal{L}_G$ only variants, local features are projected using the global projection heads.
We also report perplexity (PPL), defined as 
$\exp\!(\frac{1}{m}\sum_{m}\text{CE}(\mathbf{S}_{\cdot m}, \bm{i}_m))$, where
CE is the cross-entropy and $\bm{i}_m$ is the
ground-truth image patch index for audio frame $m$.

\textbf{Point-and-retrieve.}
To assess generalization to full-score retrieval, we introduce a point-and-retrieve task that more closely mirrors real-world use cases, e.g., a user clicking on a measure of a score image to jump to a location in a corresponding audio recording. Given a query (either an audio frame or an image patch) the model must identify its corresponding counterpart 
across \emph{all} image patches or audio frames in the complete piece. Concretely, in the A2I direction, for each audio frame we compute cosine similarity to every image patch in the full score and predict the top-1 match; symmetrically for I2A. 
We apply a tolerance of $\pm1$ image patch vertically for A2I.

\subsection{Batch Construction Strategies}
\label{subsec:batch}

A key design choice in contrastive learning is how training batches are constructed, as this determines the ``hardness'' of the negatives the model encounters during training \cite{yuksekgonul2022and}. 
We evaluate four strategies:
\textbf{(1)~Random:} Image–audio pairs are sampled uniformly at random across the full dataset.
\textbf{(2)~Same piece:} Pairs within a batch are sampled from the same musical piece, making the task harder: the model must distinguish between segments that may share harmonic or melodic content.
\textbf{(3)~Negative Mutations:}
Following~\cite{yuksekgonul2022and}, we introduce hard negatives by constructing \emph{mutated} versions of each pair.
Specifically, we pitch-shift individual notes in the source MusicXML file
by a random semitone in the range $[-4, 4]$, with each note mutated independently with probability $0.15$. The resulting mutated image–audio pairs are added to the batch and treated as negatives only.
Note that to keep total dataset size fixed across settings, we use half the number of original pairs, with the remaining half consisting of mutated pairs.
\textbf{(4)~Positive Mutations:} 
We use the same mutated pairs as above, but now treat each mutated image–audio pair as its own independent positive example rather than a hard negative. 
Compared to the  Negative Mutations strategy, this preserves the same hard negative signal while providing additional positive training pairs.
\section{Results}

\begin{figure}[t!]
    \centering
    \includegraphics[width=\linewidth]{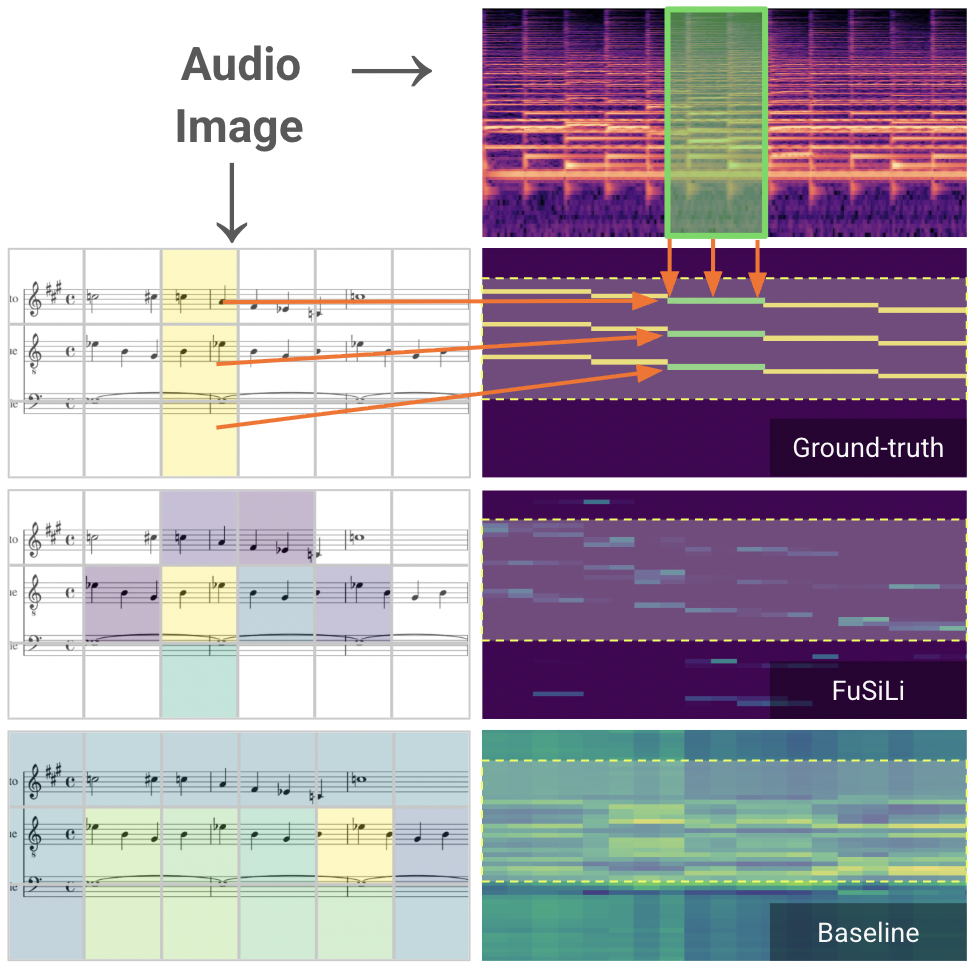}
    \vspace{-2em}
    \caption{Visualization of multimodal frame-level similarities derived from ground truth note-level labels, FuSiLi ($\alpha = 0.5$), and a global contrastive baseline ($\alpha = 0.0$).}
    \label{fig:fig2}
    \vspace{-1.5em}
\end{figure}

\begin{table*}[h!]
  \centering
  \footnotesize
  \setlength{\tabcolsep}{3pt} 
  \begin{tabular}{@{} ll
      S[table-format=1.2] S[table-format=3.2]   
      S[table-format=1.2] S[table-format=1.2]   
      S[table-format=1.2] S[table-format=1.2]   
      S[table-format=1.2] S[table-format=1.2]   
      S[table-format=1.2] S[table-format=1.2]   
      S[table-format=1.2] S[table-format=1.2]   
      S[table-format=1.2] S[table-format=1.2]   
      S[table-format=1.2] S[table-format=1.2]   
    @{}}

  \toprule
  & & \multicolumn{2}{c}{Local Eval} & \multicolumn{2}{c}{Point \& Retrieve} & \multicolumn{4}{c}{Global Retrieval (PDMX)} & \multicolumn{4}{c}{Global Retrieval (YTSV)} & \multicolumn{4}{c}{Global Retrieval (MSMD)} \\
  \cmidrule(lr){3-4} \cmidrule(lr){5-6} \cmidrule(lr){7-10} \cmidrule(lr){11-14} \cmidrule(lr){15-18}
  
  & & \multicolumn{2}{c}{A2I} & {I2A} & {A2I} & \multicolumn{2}{c}{I2A} & \multicolumn{2}{c}{A2I} & \multicolumn{2}{c}{I2A} & \multicolumn{2}{c}{A2I} & \multicolumn{2}{c}{I2A} & \multicolumn{2}{c}{A2I} \\
  \cmidrule(lr){3-4} \cmidrule(lr){5-5} \cmidrule(lr){6-6} \cmidrule(lr){7-8} \cmidrule(lr){9-10} \cmidrule(lr){11-12} \cmidrule(lr){13-14} \cmidrule(lr){15-16} \cmidrule(lr){17-18}

  {Batch} & {} & {Top-1} & {PPL} & {Top-1} & {Top-1} & {R@1} & {MRR} & {R@1} & {MRR} & {R@1} & {MRR} & {R@1} & {MRR} & {R@1} & {MRR} & {R@1} & {MRR} \\
  \midrule

  \multirow{4}{*}{Random}
    & Baseline
      & 0.19 & 35.08 & 0.01 & 0.05 & 0.75 & 0.85 & 0.75 & 0.85 & 0.01 & 0.05 & 0.02 & 0.05 & 0.19 & 0.29 & 0.12 & 0.22 \\
    & {+FuSiLi}
      & 0.24 & 35.51 & 0.10 & 0.11 & 0.72 & 0.83 & 0.72 & 0.83 & 0.01 & 0.04 & 0.01 & 0.05 & 0.06 & 0.15 & 0.10 & 0.20 \\
    & {+Neg. Mutations}
      & 0.26 & 36.03 & 0.09 & 0.09 & 0.68 & 0.79 & 0.65 & 0.76 & 0.02 & 0.05 & 0.02 & 0.05 & 0.13 & 0.24 & 0.16 & 0.28 \\
    & {+Pos. Mutations}
      & 0.17 & 37.73 & 0.09 & 0.08 & 0.67 & 0.79 & 0.67 & 0.79 & 0.02 & 0.06 & 0.01 & 0.04 & 0.17 & 0.29 & 0.24 & 0.35 \\
  \midrule

  \multirow{4}{*}{\makecell[l]{Same\\Piece}}
    & Baseline
      & 0.16 & 38.11 & 0.01 & 0.05 & 0.77 & 0.86 & 0.76 & 0.86 & 0.04 & 0.11 & 0.04 & 0.09 & 0.22 & 0.34 & 0.24 & 0.37 \\
    & {+FuSiLi}
      & 0.21 & 33.86 & 0.19 & 0.11 & 0.76 & 0.86 & 0.76 & 0.86 & 0.05 & 0.11 & 0.03 & 0.08 & 0.27 & 0.37 & 0.27 & 0.39 \\
    & {+Neg. Mutations}
      & 0.27 & 34.39 & 0.16 & 0.12 & 0.71 & 0.82 & 0.72 & 0.83 & 0.02 & 0.07 & 0.03 & 0.08 & 0.31 & 0.43 & 0.29 & 0.42 \\
    & {+Pos. Mutations}
      & 0.30 & 34.24 & 0.14 & 0.14 & 0.72 & 0.83 & 0.73 & 0.83 & 0.05 & 0.10 & 0.04 & 0.08 & 0.25 & 0.38 & 0.25 & 0.39 \\

  \midrule
  & U-MusT~\cite{jung2025unified}
  & 0.20 & 101.27 & 0.23 & 0.12 & {---} & {---} & {---} & {---} & {---} & {---} & {---} & {---} & {---} & {---} & {---} & {---} \\
  \bottomrule
  \end{tabular}
  \vspace{-0.5em}
  \caption{Comparison of global baseline and FuSiLi across batch construction strategies. 
  Results are reported for in-domain (PDMX/MSMD) and out-of-domain (YTSV) global retrieval, local alignment (MSMD), and point-and-retrieve tasks.}
\label{tab:results_comprehensive}
\end{table*}

\begin{table*}[t]
  \centering
  \footnotesize
  \setlength{\tabcolsep}{3pt} 
  \begin{tabular}{@{} ll 
      S[table-format=1.2] S[table-format=2.2] 
      S[table-format=1.2] S[table-format=1.2] 
      S[table-format=1.2] S[table-format=1.2] 
      S[table-format=1.2] S[table-format=1.2] 
      S[table-format=1.2] S[table-format=1.2] 
      S[table-format=1.2] S[table-format=1.2] 
      S[table-format=1.2] S[table-format=1.2] 
      S[table-format=1.2] S[table-format=1.2] 
    @{}}
  \toprule
  & & \multicolumn{2}{c}{Local Eval} & \multicolumn{2}{c}{Point \& Retrieve} & \multicolumn{4}{c}{Global Retrieval (PDMX)} & \multicolumn{4}{c}{Global Retrieval (YTSV)} & \multicolumn{4}{c}{Global Retrieval (MSMD)} \\
  \cmidrule(lr){3-4} \cmidrule(lr){5-6} \cmidrule(lr){7-10} \cmidrule(lr){11-14} \cmidrule(lr){15-18}

  & & \multicolumn{2}{c}{A2I} & \multicolumn{1}{c}{I2A} & \multicolumn{1}{c}{A2I} & \multicolumn{2}{c}{I2A} & \multicolumn{2}{c}{A2I} & \multicolumn{2}{c}{I2A} & \multicolumn{2}{c}{A2I} & \multicolumn{2}{c}{I2A} & \multicolumn{2}{c}{A2I} \\
  \cmidrule(lr){3-4} \cmidrule(lr){5-5} \cmidrule(lr){6-6} \cmidrule(lr){7-8} \cmidrule(lr){9-10} \cmidrule(lr){11-12} \cmidrule(lr){13-14} \cmidrule(lr){15-16} \cmidrule(lr){17-18}

  {$\alpha$} & {Local Sim.} & {Top-1} & {PPL} & {Top-1} & {Top-1} & {R@1} & {MRR} & {R@1} & {MRR} & {R@1} & {MRR} & {R@1} & {MRR} & {R@1} & {MRR} & {R@1} & {MRR} \\
  \midrule

  \rowcolor{ours}
  {$0.5 (\mathcal{L}_G)$} & {FuSiLi $(\mathcal{L}_L)$}
    & 0.30 & 34.24 & 0.14 & 0.14 & 0.72 & 0.83 & 0.73 & 0.83 & 0.05 & 0.10 & 0.04 & 0.08 & 0.25 & 0.38 & 0.25 & 0.39 \\

  {$0.5 (\mathcal{L}_G)$} & {Cosine $(\mathcal{L}_\text{cos})$}
    & 0.02 & 42.49 & 0.02 & 0.02 & 0.51 & 0.66 & 0.51 & 0.67 & 0.00 & 0.02 & 0.00 & 0.02 & 0.05 & 0.11 & 0.06 & 0.14 \\

  {$0 (\mathcal{L}_G)$} & {$\times$}
    & 0.24 & 36.55 & 0.01 & 0.05 & 0.73 & 0.83 & 0.73 & 0.83 & 0.07 & 0.12 & 0.03 & 0.07 & 0.27 & 0.38 & 0.28 & 0.41 \\
  
  {$1$} & {FuSiLi $(\mathcal{L}_L)$}
    & 0.16 & 32.41 & 0.16 & 0.08 & 0.27 & 0.38 & 0.13 & 0.24 & 0.00 & 0.03 & 0.01 & 0.02 & 0.09 & 0.16 & 0.06 & 0.13 \\

  {$1$} & {Cosine $(\mathcal{L}_\text{cos})$}
    & 0.03 & 42.46 & 0.01 & 0.01 & 0.00 & 0.02 & 0.00 & 0.02 & 0.01 & 0.02 & 0.00 & 0.01 & 0.00 & 0.02 & 0.00 & 0.02 \\
  \bottomrule
  \end{tabular}
  \vspace{-0.5em}
  \caption{Effect of training objective and local similarity formulation under the Same Piece + Pos. Mutations batch setting. We vary the global/local loss balance $\alpha$ and compare the proposed FuSiLi similarity against a cosine-based alternative. Results are reported on global retrieval, local alignment, and point-and-retrieve tasks.}
  \label{tab:results_ablations}
  \vspace{-1em}
\end{table*}

\label{sec:results}

\subsection{FuSiLi vs. Global Baseline}
\label{subsec:main_results}

We first assess how FuSiLi compares against the global baseline and seq2seq model across different batch construction strategies, with results 
in Table~\ref{tab:results_comprehensive}.

\textbf{FuSiLi consistently improves local alignment over the global baseline.} Across both batch designs, using FuSiLi as the similarity score in training yields clear gains in frame-level alignment. The top-1 accuracy for the best performance FuSiLi variant (Same Piece + Pos. Mutations) on local alignment is nearly twice the Same Piece baseline. Additionally, perplexity drops 4 points from the baseline to the same FuSiLi variant, reflecting higher confidence in local alignment. Point\&Retrieve results also show substantial improvement: in the I2A direction, FuSiLi variants achieve over $10\times$ improvement over the global baseline ($13.91\%$ vs $1.33\%$) and nearly $3\times$ in the A2I direction ($14\%$ vs. $5\%$). These results confirm that FuSiLi is successful at learning frame-level correspondences between image and audio pairs from global supervision alone; see Figure~\ref{fig:fig2} for a qualitative illustration.

\textbf{FuSiLi preserves global retrieval performance.} Despite incorporating a local similarity score, FuSiLi variants trained with the hybrid objective ($\alpha=0.5$) do not sacrifice retrieval ability as the global loss term successfully supervises the pooled embeddings. On PDMX and MSMD, these variants remain competitive with the global baseline. 
On the out-of-domain YTSV set consisting of real-world score scans and audio recordings, FuSiLi variants remain on par with the global baseline, suggesting that the domain shift affects both approaches equally.

\textbf{Batch design matters.} Same Piece batches consistently outperform Random batches on local alignment tasks, suggesting that constructing a batch with more challenging negatives forces the model to learn fine-grained distinctions. Among Same Piece variants, FuSiLi + Pos. Mutations achieves the best local alignment top-1 accuracy and point-and retrieve A2I top-1 accuracy, while just FuSiLi achieves the strongest point-and-retrieve performance in the I2A direction. However, the mutation-based variants introduce a trade-off: because total dataset size is fixed
(see Section~\ref{subsec:datasets}), half of all training pairs are near-duplicates of the other half. This reduction in diversity may explain why mutation variants sometimes underperform simpler strategies, particularly in the Random setting, and suggests that expanding the dataset with mutations rather than replacing original pairs could yield further gains.

\textbf{FuSiLi outperforms seq2seq in most settings at lower inference cost.}
U-MuST achieves $20\%$ top-1 local alignment accuracy, which outperforms the global baseline but falls short of our best FuSiLi variant. It also has a notably higher perplexity than all contrastive models, indicating that while U-MuST can identify the correct image patch with reasonable top-1 accuracy, its alignment distributions are considerably less confident than those of FuSiLi. On point-and-retrieve, U-MuST outperforms our best variants but shows worse performance on A2I. Notably, U-MuST requires substantially more compute at inference time and a more complex preprocessing pipeline, whereas FuSiLi achieves competitive or superior alignment using only efficient embedding-based inference.

\subsection{Effects of $\alpha$ and Local Similarity Score}
\label{subsec:ablation_results}

Fixing the batch configuration to Same Piece + Pos.\ Mutations, we investigate the effect of the balancing hyperparameter $\alpha$ and the choice of local similarity score formulation. We compare three values of $\alpha \in \{0.0, 0.5, 1.0\}$, corresponding to training with $\mathcal{L}_G$ only, the hybrid objective, and $\mathcal{L}_L$ only respectively. For configurations that include a local loss, we compare our proposed FuSiLi score $\batchsim_{ij}^{L}$ against a cosine-based alternative $\batchsim_{ij}^{\text{cos}}$, defined as the sum of all entries of the \atom-wise cosine similarity matrix between local features, with induced loss $\mathcal{L}_\text{cos}$ (Table~\ref{tab:results_ablations}).

\textbf{The Sinkhorn-based similarity score is essential.}
Replacing FuSiLi with a summed cosine similarity score ($\mathcal{L}_\text{cos}$) leads to a significant drop across all metrics. In the hybrid setting ($\mathcal{L}_G + \mathcal{L}_\text{cos}$), local alignment top-1 accuracy falls to $2\%$ and retrieval collapses on YTSV, suggesting that the cosine objective fails to improve local alignment \emph{and} negatively affects global retrieval. We hypothesize that the cosine local objective either conflicts with the global objective or
fails to provide useful supervision while reducing the relative contribution of the global objective.

\textbf{The hybrid objective balances local and global performance.}
Training with $\mathcal{L}_L$ alone achieves the strongest point-and-retrieve I2A performance ($16\%$) but degrades retrieval substantially.
Adding $\mathcal{L}_G$ recovers retrieval performance while maintaining strong local alignment, with ($\alpha = 0.5$)
representing the best overall trade-off.

\section{Conclusion}

We propose FuSiLi, a similarity score function for contrastive learning that induces frame-level alignments between score images and audio from global supervision alone, using a fuse-then-pool approach grounded in Sinkhorn-based soft alignment to learn local correspondences. Our experiments show large improvements in local alignment and zero-shot point-and-retrieve tasks over standard contrastive baselines while preserving competitive global retrieval performance. 
FuSiLi's ability to learn local information from global supervision is promising for other domains with scarce fine-grained annotations, 
and opens avenues for work on stronger alignment inductive biases and scaling to larger models or paired datasets.
\section{Acknowledgments}

This work was supported by the Humanities and AI Virtual Institute (HAVI) at Schmidt Sciences.
\section{Ethics Statement}

While our work sits within the fraught landscape of music and AI, we do not believe our specific focus here carries with it any noteworthy ethical risks. 
Our efforts to improve multimodal music understanding are motivated by bolstering improvements in broad tasks that may aid human musicians in every day creative workflows, e.g.,~point-and-retrieve, score following, automatic music transcription, and optical music recognition.

\bibliography{references.bib}

@inproceedings{dannenberg1984line,
  title={An on-line algorithm for real-time accompaniment},
  author={Dannenberg, Roger B},
  booktitle={ICMC},
  volume={84},
  pages={193--198},
  year={1984}
}

@inproceedings{dixon2005line,
  title={An On-Line Time Warping Algorithm for Tracking Musical Performances.},
  author={Dixon, Simon},
  booktitle={IJCAI},
  pages={1727--1728},
  year={2005}
}

@inproceedings{cont2006realtime,
  title={Realtime audio to score alignment for polyphonic music instruments, using sparse non-negative constraints and hierarchical HMMs},
  author={Cont, Arshia},
  booktitle={IEEE International Conference on Acoustics Speech and Signal Processing (ICASSP)},
  year={2006}
}

@inproceedings{dorfer2017learning,
  title={Learning audio-sheet music correspondences for score identification and offline alignment},
  author={Dorfer, Matthias and Arzt, Andreas and Widmer, Gerhard},
  booktitle={Proceedings of the 18th International Society for Music Information Retrieval Conference (ISMIR)},
  year={2017}
}

@inproceedings{balke2019learning,
  title={Learning soft-attention models for tempo-invariant audio-sheet music retrieval},
  author={Balke, Stefan and Dorfer, Matthias and Carvalho, Luis and Arzt, Andreas and Widmer, Gerhard},
  booktitle={Proceedings of the 20th International Society for Music Information Retrieval Conference (ISMIR)},
  year={2019}
}

@article{henkel2021real,
  title={Real-time music following in score sheet images via multi-resolution prediction},
  author={Henkel, Florian and Widmer, Gerhard},
  journal={Frontiers in Computer Science},
  volume={3},
  pages={718340},
  year={2021}
}

@inproceedings{bukey2024just,
  title={Just label the repeats for in-the-wild audio-to-score alignment},
  author={Bukey, Irmak and Feffer, Michael and Donahue, Chris},
  booktitle={Proceedings of the 25th International Society for Music Information Retrieval Conference (ISMIR)},
  year={2024}
}

@inproceedings{carvalho2023towards,
  title={Towards robust and truly large-scale audio-sheet music retrieval},
  author={Carvalho, Lu{\'\i}s and Widmer, Gerhard},
  booktitle={IEEE 6th International Conference on Multimedia Information Processing and Retrieval (MIPR)},
  year={2023}
}

@inproceedings{carvalho2023self,
  title={Self-supervised contrastive learning for robust audio-sheet music retrieval systems},
  author={Carvalho, Lu{\'\i}s and Wash{\"u}ttl, Tobias and Widmer, Gerhard},
  booktitle={Proceedings of the 14th Conference on ACM Multimedia Systems},
  year={2023}
}

@inproceedings{radford2021learning,
  title={Learning transferable visual models from natural language supervision},
  author={Radford, Alec and Kim, Jong Wook and Hallacy, Chris and Ramesh, Aditya and Goh, Gabriel and Agarwal, Sandhini and Sastry, Girish and Askell, Amanda and Mishkin, Pamela and Clark, Jack and others},
  booktitle={International Conference on Machine Learning (ICML)},
  year={2021}
}

@article{oord2018representation,
  title={Representation learning with contrastive predictive coding},
  author={Oord, Aaron van den and Li, Yazhe and Vinyals, Oriol},
  journal={arXiv preprint arXiv:1807.03748},
  year={2018}
}

@inproceedings{elizalde2023clap,
  title={Clap learning audio concepts from natural language supervision},
  author={Elizalde, Benjamin and Deshmukh, Soham and Al Ismail, Mahmoud and Wang, Huaming},
  booktitle={IEEE International Conference on Acoustics, Speech and Signal Processing (ICASSP)},
  year={2023},
}

@inproceedings{wu2023large,
  title={Large-scale contrastive language-audio pretraining with feature fusion and keyword-to-caption augmentation},
  author={Wu, Yusong and Chen, Ke and Zhang, Tianyu and Hui, Yuchen and Berg-Kirkpatrick, Taylor and Dubnov, Shlomo},
  booktitle={IEEE International Conference on Acoustics, Speech and Signal Processing (ICASSP)},
  year={2023},
}

@inproceedings{manco2022contrastive,
  title={Contrastive audio-language learning for music},
  author={Manco, Ilaria and Benetos, Emmanouil and Quinton, Elio and Fazekas, Gy{\"o}rgy},
  booktitle={Proceedings of the 23rd International Society for Music Information Retrieval Conference (ISMIR)},
  year={2022}
}

@inproceedings{wu2025collap,
  title={Collap: Contrastive long-form language-audio pretraining with musical temporal structure augmentation},
  author={Wu, Junda and Li, Warren and Novack, Zachary and Namburi, Amit and Chen, Carol and McAuley, Julian},
  booktitle={IEEE International Conference on Acoustics, Speech and Signal Processing (ICASSP)},
  year={2025}
}

@inproceedings{wu2025flam,
  title={Flam: Frame-wise language-audio modeling},
  author={Wu, Yusong and Tsirigotis, Christos and Chen, Ke and Huang, Cheng-Zhi Anna and Courville, Aaron and Nieto, Oriol and Seetharaman, Prem and Salamon, Justin},
  booktitle={International Conference on Machine Learning (ICML)},
  year={2025}
}

@inproceedings{li2026finelap,
  title={FineLAP: Taming Heterogeneous Supervision for Fine-grained Language-Audio Pretraining},
  author={Li, Xiquan and Xu, Xuenan and Ma, Ziyang and Chen, Wenxi and He, Haolin and Kong, Qiuqiang and Chen, Xie},
  booktitle={Proceedings of the 64th Annual Meeting of the Association for Computational Linguistics},
  year={2026}
}

@article{cuturi2013sinkhorn,
  title={Sinkhorn distances: Lightspeed computation of optimal transport},
  author={Cuturi, Marco},
  journal={Advances in Neural Information Processing Systems},
  volume={26},
  year={2013}
}

@article{chen2024your,
  title={Your contrastive learning problem is secretly a distribution alignment problem},
  author={Chen, Zihao and Lin, Chi-Heng and Liu, Ran and Xiao, Jingyun and Dyer, Eva L},
  journal={Advances in Neural Information Processing Systems},
  volume={37},
  pages={91597--91617},
  year={2024}
}

@inproceedings{wu2025clamp,
  title={Clamp 3: Universal music information retrieval across unaligned modalities and unseen languages},
  author={Wu, Shangda and Zhancheng, Guo and Yuan, Ruibin and Jiang, Junyan and Doh, Seungheon and Xia, Gus and Nam, Juhan and Li, Xiaobing and Yu, Feng and Sun, Maosong},
  booktitle={Findings of the Association for Computational Linguistics (ACL)},
  pages={2605--2625},
  year={2025}
}

@inproceedings{primus2025tacos,
  title={Tacos: Temporally-aligned audio captions for language-audio pretraining},
  author={Primus, Paul and Schmid, Florian and Widmer, Gerhard},
  booktitle={IEEE Workshop on Applications of Signal Processing to Audio and Acoustics (WASPAA)},
  year={2025}
}

@inproceedings{li2024advancing,
  title={Advancing multi-grained alignment for contrastive language-audio pre-training},
  author={Li, Yiming and Guo, Zhifang and Wang, Xiangdong and Liu, Hong},
  booktitle={Proceedings of the 32nd ACM International Conference on Multimedia},
  pages={7356--7365},
  year={2024}
}

@inproceedings{chen2024contrastive,
  title={Contrastive localized language-image pre-training},
  author={Chen, Hong-You and Lai, Zhengfeng and Zhang, Haotian and Wang, Xinze and Eichner, Marcin and You, Keen and Cao, Meng and Zhang, Bowen and Yang, Yinfei and Gan, Zhe},
  booktitle={International Conference on Machine Learning (ICML)},
  year={2025}
}

@inproceedings{gandelsman2023interpreting,
  title={Interpreting clip's image representation via text-based decomposition},
  author={Gandelsman, Yossi and Efros, Alexei A and Steinhardt, Jacob},
  booktitle={International Conference on Learning Representations (ICLR)},
  year={2024}
}

@inproceedings{bousselham2024grounding,
  title={Grounding everything: Emerging localization properties in vision-language transformers},
  author={Bousselham, Walid and Petersen, Felix and Ferrari, Vittorio and Kuehne, Hilde},
  booktitle={Proceedings of the IEEE/CVF Conference on Computer Vision and Pattern Recognition},
  year={2024}
}

@inproceedings{emon2025whisq,
  title={Whisq: Cross-modal representation learning for text-to-music mos prediction},
  author={Emon, Jakaria Islam and Salek, Md Abu and Alam, Kazi Tamanna},
  booktitle={IEEE Automatic Speech Recognition and Understanding Workshop (ASRU)},
  year={2025},
}

@article{caron2020unsupervised,
  title={Unsupervised learning of visual features by contrasting cluster assignments},
  author={Caron, Mathilde and Misra, Ishan and Mairal, Julien and Goyal, Priya and Bojanowski, Piotr and Joulin, Armand},
  journal={Advances in Neural Information Processing Systems},
  volume={33},
  pages={9912--9924},
  year={2020}
}

@article{jung2025unified,
  title={U-MusT: A Unified Framework for Cross-modal Translation of Score Images, Symbolic Music, and Performance Audio},
  author={Jung, Jongmin and Kim, Dongmin and Lee, Sihun and Cho, Seola and So, Hyungjoon and Bukey, Irmak and Donahue, Chris and Jeong, Dasaem},
  journal={IEEE Transactions on Audio, Speech and Language Processing},
  year={2025}
}

@inproceedings{feffer2022assistive,
  title={Assistive alignment of in-the-wild sheet music and performances},
  author={Feffer, Michael and Donahue, Chris and Lipton, Zachary},
  booktitle={Proceedings of the 23rd International Society for Music Information Retrieval Conference (ISMIR)},
  year={2022}
}

@software{ilharco_gabriel_2021_5143773,
  author       = {Ilharco, Gabriel and
                  Wortsman, Mitchell and
                  Wightman, Ross and
                  Gordon, Cade and
                  Carlini, Nicholas and
                  Taori, Rohan and
                  Dave, Achal and
                  Shankar, Vaishaal and
                  Namkoong, Hongseok and
                  Miller, John and
                  Hajishirzi, Hannaneh and
                  Farhadi, Ali and
                  Schmidt, Ludwig},
  title        = {OpenCLIP},
  month        = jul,
  year         = 2021,
  publisher    = {Zenodo},
  version      = {0.1},
  doi          = {10.5281/zenodo.5143773},
  url          = {https://doi.org/10.5281/zenodo.5143773}
}

@inproceedings{htsatke2022,
  author = {Ke Chen and Xingjian Du and Bilei Zhu and Zejun Ma and Taylor Berg-Kirkpatrick and Shlomo Dubnov},
  title = {HTS-AT: A Hierarchical Token-Semantic Audio Transformer for Sound Classification and Detection},
  booktitle={IEEE International Conference on Acoustics, Speech and Signal Processing (ICASSP)},
  year = {2022}
}

@inproceedings{yuksekgonul2022and,
  title={When and why vision-language models behave like bags-of-words, and what to do about it?},
  author={Yuksekgonul, Mert and Bianchi, Federico and Kalluri, Pratyusha and Jurafsky, Dan and Zou, James},
  booktitle={International Conference on Learning Representations (ICLR)},
  year={2023}
}

@article{dorfer2018learning,
  title={Learning audio--sheet music correspondences for cross-modal retrieval and piece identification},
  author={Dorfer, Matthias and Haji{\v{c}} Jr, Jan and Arzt, Andreas and Frostel, Harald and Widmer, Gerhard},
  journal={Transactions of the International Society for Music Information Retrieval (TISMIR)},
  volume={1},
  number={1},
  year={2018}
}

@inproceedings{long2025pdmx,
  title={Pdmx: A large-scale public domain musicxml dataset for symbolic music processing},
  author={Long, Phillip and Novack, Zachary and Berg-Kirkpatrick, Taylor and McAuley, Julian},
  booktitle={IEEE International Conference on Acoustics, Speech and Signal Processing (ICASSP)},
  year={2025}
}

@inproceedings{vercoe1984synthetic,
  title={The synthetic performer in the context of live performance},
  author={Vercoe, Barry},
  booktitle={Proceedings of International Computer Music Conference},
  pages={199--200},
  year={1984}
}

@article{zeng2021contrastive,
  title={Contrastive learning of global and local video representations},
  author={Zeng, Zhaoyang and McDuff, Daniel and Song, Yale and others},
  journal={Advances in Neural Information Processing Systems},
  volume={34},
  pages={7025--7040},
  year={2021}
}

@article{anonto2025align,
  title={Align Where the Words Look: Cross-Attention-Guided Patch Alignment with Contrastive and Transport Regularization for Bengali Captioning},
  author={Anonto, Riad Ahmed and Zabin, Sardar Md Saffat and Rahman, M Saifur},
  journal={arXiv preprint arXiv:2509.18369},
  year={2025}
}

@inproceedings{cuturi2017soft,
  title={Soft-dtw: a differentiable loss function for time-series},
  author={Cuturi, Marco and Blondel, Mathieu},
  booktitle={International Conference on Machine Learning (ICML)},
  year={2017}
}

@article{benetos2018automatic,
  title={Automatic music transcription: An overview},
  author={Benetos, Emmanouil and Dixon, Simon and Duan, Zhiyao and Ewert, Sebastian},
  journal={IEEE Signal Processing Magazine},
  volume={36},
  number={1},
  pages={20--30},
  year={2018},
  publisher={IEEE}
}

@article{calvo2020understanding,
  title={Understanding optical music recognition},
  author={Calvo-Zaragoza, Jorge and Jr, Jan Haji{\v{c}} and Pacha, Alexander},
  journal={ACM Computing Surveys (CSUR)},
  volume={53},
  number={4},
  pages={1--35},
  year={2020},
  publisher={ACM New York, NY, USA}
}

@article{orio2003score,
  title={Score following: State of the art and new developments},
  author={Orio, Nicola and Lemouton, Serge and Schwarz, Diemo},
  journal={New Interfaces for Musical Expression (NIME)},
  year={2003}
}

\clearpage
\appendix
\section{Supplementary Material}

\subsection{Implementation Details}

\subsubsection{Training Setup}
All models share the same base training setup regardless of objective. Encoder parameters are updated with a learning rate of $2 \times 10^{-5}$, while the projection layers use a higher learning rate of $2 \times 10^{-4}$. We use the Adam optimizer with weight decay $0.01$. Training uses a batch size of 128, distributed across 4 GPUs (32 samples per GPU), for up to 20 epochs with early stopping based on a validation criterion described below.

\subsubsection{Audio Preprocessing}
Since the CLAP encoder natively accepts 10-second inputs, we split each 20-second audio clip into two 10-second segments, encode them separately, and concatenate the resulting frame embeddings. This yields a sequence of 2048 audio frames, which we subsample by a factor of 8 via max pooling, resulting in 256 audio frames per clip.

\subsubsection{Sinkhorn Hyperparameters and Model Selection}
For model variants that include $\mathcal{L}_L$, the Sinkhorn algorithm uses $K = 20$ iterations with temperature $\epsilon = 0.07$, subsequently learned during training. 

 During training, we track two validation metrics: (1) the validation loss under the model's training objective (e.g., $\alpha \mathcal{L}_L + (1 - \alpha)\mathcal{L}_G$), and (2) top-1 frame-level alignment accuracy on the MSMD validation set. We select the checkpoint that achieves the best joint performance across both metrics.

\subsection{Evaluation Details for the Seq2Seq Baseline}
Due to differences in image resolution between U-MusT and our contrastive models, we apply a relaxed matching criterion for U-MusT 
across all tasks:
a predicted image patch is considered correct if the ground-truth patch falls within a neighborhood of $\pm2$ patches vertically or $\pm1$ patch horizontally. Additionally, 
we exclude patches corresponding to the first three columns---which typically contain clef, key, and time signature markings---as the model consistently assigns disproportionally high attention weight to these patches, obscuring the model's true alignment capability.
For the point-and-retrieve task specifically, in the A2I direction we extend the vertical tolerance to the full image column, making it comparable to the $\pm1$ patch tolerance applied to our contrastive models.

\subsection{Experimental Configurations}

\subsubsection{Experiment 1: Global Baseline vs. FuSiLi across batch construction strategies}
\label{subsec:exp1}
We evaluate eight models in addition to the seq2seq baseline, organized as four variants across two batch designs (Random and Same Piece): (1) \textbf{Baseline}, model trained with $\mathcal{L}_G$ only ($\alpha = 0$); (2) \textbf{+FuSiLi}, model trained with the hybrid objective $\mathcal{L}$ with $\alpha = 0.5$; (3) \textbf{+Neg.\ Mutations}, FuSiLi with mutated pairs added to the batch as hard negatives; and (4) \textbf{+Pos.\ Mutations}, FuSiLi with mutated pairs added as additional self-consistent positive examples.

\subsubsection{Experiment 2: Training Objective and Similarity Score}
\label{subsec:exp2}
Fixing the batch configuration to Same Piece + Pos.\ Mutations, we investigate the effect of the balancing hyperparameter $\alpha$ and the choice of local similarity score formulation. We compare three values of $\alpha \in \{0.0, 0.5, 1.0\}$, corresponding to training with $\mathcal{L}_G$ only, the hybrid objective, and $\mathcal{L}_L$ only respectively. For configurations that include a local loss, we compare our proposed FuSiLi score $\batchsim_{ij}^{L}$ against a cosine-based alternative $\batchsim_{ij}^{\text{cos}}$, defined as the sum of all entries of the \atom-wise cosine similarity matrix between local features, with a corresponding InfoNCE loss $\mathcal{L}_\text{cos}$.
Note that $\batchsim_{ij}^{\text{cos}}$ is still computed on full 2D local representations, serving as a direct baseline for the proposed FuSiLi score. 

\end{document}